\documentclass[review]{elsarticle}

\usepackage{lineno}
\usepackage{amsmath}
\usepackage{graphicx}
\usepackage{url}
\usepackage{epstopdf}
\journal{Journal of \LaTeX\ Templates}

\begin{document}

\begin{frontmatter}

\title{Application of the Compress Sensing Theory for Improvement of the TOF Resolution in a Novel J-PET Instrument}

\author[SWIERK]{L.~Raczy\'nski}
\author[WFAIS]{P.~Moskal}
\author[SWIERK]{P.~Kowalski}
\author[SWIERK]{W.~Wi\'slicki}
\author[WFAIS]{T.~Bednarski}
\author[WFAIS]{P.~Bia\l as}
\author[WFAIS]{E.~Czerwi\'nski}
\author[WFAIS]{A.~Gajos}
\author[WFAIS,PAN]{\L .~Kap\l on}
\author[WCHUJ]{A.~Kochanowski}
\author[WFAIS]{G.~Korcyl} 
\author[WFAIS]{J.~Kowal}
\author[WFAIS]{T.~Kozik}
\author[WFAIS]{W.~Krzemie\'n}
\author[WFAIS]{E.~Kubicz}
\author[WFAIS]{Sz.~Nied\'zwiecki}
\author[WFAIS]{M.~Pa\l ka}
\author[WFAIS]{Z.~Rudy}
\author[WFAIS]{P.~Salabura}
\author[WFAIS]{N.G.~Sharma}
\author[WFAIS]{M.~Silarski}
\author[WFAIS]{A.~S\l omski} 
\author[WFAIS]{J.~Smyrski}
\author[WFAIS]{A.~Strzelecki}
\author[WFAIS,PAN]{A.~Wieczorek}
\author[WFAIS]{M.~Zieli\'nski}
\author[WFAIS]{N.~Zo\'n}

\address[SWIERK]{\'Swierk Computing Centre, National Centre for Nuclear Research, 05-400 Otwock-\'Swierk, Poland}
\address[WFAIS]{Faculty of Physics, Astronomy and Applied Computer Science,
 Jagiellonian University, 30-059 Cracow, Poland}
\address[PAN]{Institute of Metallurgy and Materials Science of Polish Academy of Sciences, Cracow, Poland}
\address[WCHUJ]{Faculty of Chemistry, Jagiellonian University, 30-060 Cracow, Poland}

\begin{abstract}
Nowadays, in Positron Emission Tomography (PET) systems, a Time of Flight information is used to improve the image reconstruction process. In Time of Flight PET (TOF-PET), fast detectors are able to measure the difference in the arrival time of the two gamma rays, with the precision enabling to shorten significantly a range along the line-of-response (LOR) where the annihilation occurred. In the new concept, called J-PET scanner, gamma rays are detected in plastic scintillators. In a single strip of J-PET system, time values are obtained by probing signals in the amplitude domain. Owing to Compress Sensing theory, information about the shape and amplitude of the signals is recovered. In this paper we demonstrate that based on the acquired signals parameters, a better signal normalization may be provided in order to improve the TOF resolution. The procedure was tested using large sample of data registered by a dedicated detection setup enabling sampling of signals with 50~ps intervals. Experimental setup provided irradiation of a chosen position in the plastic scintillator strip with annihilation gamma quanta.

\end{abstract}

\begin{keyword}
\texttt{Compressed Sensing} \sep \texttt{Positron Emission Tomography} \sep \texttt{Time-of-Flight}
%\MSC[2010] 00-01\sep  99-00
\end{keyword}

\end{frontmatter}

\section{INTRODUCTION}

Positron Emission Tomography (PET) [1] is currently one of the most perspective techniques in the field of medical imaging. PET is based on the fact that the electron and positron annihilate and their mass is converted to energy in the form of two gamma quanta flying in the opposite directions. The two gamma quanta registered in coincidence define a line referred to as line of response (LOR). The image of distribution of the radionuclide is obtained from the high statistics sample of reconstructed LORs. In Time of Flight PET (TOF-PET) systems [2, 3], the applied detectors measure the difference in the arrival time of the two gamma rays which enables to shorten significantly a range along the LOR used for the reconstruction of the image.

Currently all commercial PET devices use inorganic scintillator materials, usually LSO or LYSO crystals, as radiation detectors. These are characterized by relatively long rise- and decay times, of the order of tens of nanoseconds. The J-PET collaboration investigates a possibility of construction of a PET scanner from plastic scintillators which would allow for simultaneous imaging of the whole human body. The J-PET chamber is built out from long strips forming the cylinder [4, 5]. Light signals from each strip are converted into electrical signals by two photomultipliers (PM) placed at opposite edges. It should be noted that the better the time resolutions of the detection system the better is the quality of a reconstructed images.

In this paper we will investigate the TOF resolution of a novel J-PET scanner and we will show a simple method to improve the results by applying the compressive sensing theory. In the following we define the time resolution and present shortly the method of signal normalization based on compressive sensing theory. Then we describe an experimental setup used for signal registration and present results of improving TOF resolution in 30~cm long plastic scintillator strip, read out on both sides by the Hamamatsu R4998 photomultipliers. Signals from the photomultipliers were sampled in 50~ps steps using the Lecroy Signal Data Analyzer 6000A.

\section{MATERIALS AND METHODS}

\subsection{Definition of time resolution}

In the following we will be interested in determination of moment of interaction of gamma quantum with a strip. The interaction moment ($t_{hit}$) is given by:
\begin{equation}
	t_{hit} = \frac{t_{1} + t_{2}}{2} - \frac{D}{2v_{eff}},	
	\label{thit}
\end{equation}
where $t_1$ and $t_2$ are the arrival times to the PM1(2), respectively, $D$ is the length of whole strip, and $v_{eff}$ is the effective speed of the light in used scintillator. In the recent work [6], the speed of the light in the scintillator was estimated to 12.6 cm/ns.

In order to determine the resolution (standard deviation) of thit determination an indirect method based on the estimation of the resolution of time difference ($\Delta t$) will be provided. We assume for the sake of simplicity that $v_{eff}$ in Eq. 1 is known exactly. Since the time difference $\Delta t = t_2 - t_1$, we have
\begin{equation}
	\sigma^2 (\Delta t) = \sigma^2(t_{1}) + \sigma^2(t_{2})	
\end{equation}
and the resolution of $t_{hit}$ based on Eq.~\ref{thit} may be expressed as:
\begin{equation}
	\sigma^2 (t_{hit}) = \frac{\sigma^2(t_{1}) + \sigma^2(t_{2})}{4} = \frac{\sigma^2 (\Delta t)}{4}	
\end{equation}
which implies that the resolution of the determination of interaction moment ($t_{hit}$) is twice better than the resolution of the time difference $\Delta t$.

\subsection{Experimental Setup}

A necessary data to carry out the research have been acquired by a single module of the J-PET detector [6]. The scheme of experimental setup is presented in Fig. 1. The 30 cm long strip was connected on two sides to the photomultipliers (PMs). The radioactive $^{22}Na$ source was moved from the first to the second end in steps of 6~mm. At each position, about 10~000 pairs of signals from PM1 and PM2 were registered in coincidence with reference detector. The signals were sampled using the scope with a probing interval of 50~ps. Examples of two signals registered at PM1 and PM2 are shown in Fig. 2 with blue (red) colors, respectively, for the case when the scintillator was irradiated at distance of 7 cm from PM2 (23 cm from PM1).
\begin{figure}[h!]
	\centerline{\includegraphics[width=0.7\textwidth]{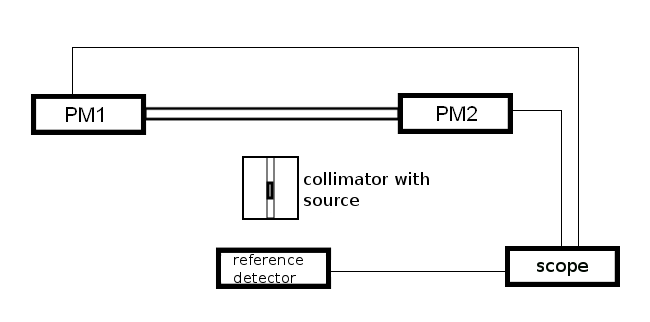}}
\caption{
Experimental setup. 
\label{Exper_setup}
}
\end{figure}

In the final, multi-modular devices with hundreds of photomultipliers probing with scopes will not be possible. Therefore, a multi-threshold sampling method to generate samples of a PET event waveform with respect to four user-defined amplitudes was proposed. An electronic system for probing these signals in a voltage domain was developed and successfully tested [7]. Based on the signals registered via scope, we simulate a four-level measurement with sampling in the voltage domain at 50, 100, 200 and 300~mV, indicated by four black horizontal lines in Fig. 2. It should be stressed that due to the time walk effect the resolution determined when applying the lowest threshold (50~mV) is better with respect to the resolution obtained at the highest level (300~mV). Therefore the simplest way to define the start of each pulse, times $t_1$ and $t_2$, is to use the information from registration time at the lowest amplitude level, marked with vertical dashed lines in Fig. 2. From this one may easily estimate the resolution of $\Delta t$ and therefore the resolution of thit (see Eq. 3). 

\begin{figure}[h!]
	\centerline{\includegraphics[width=0.7\textwidth]{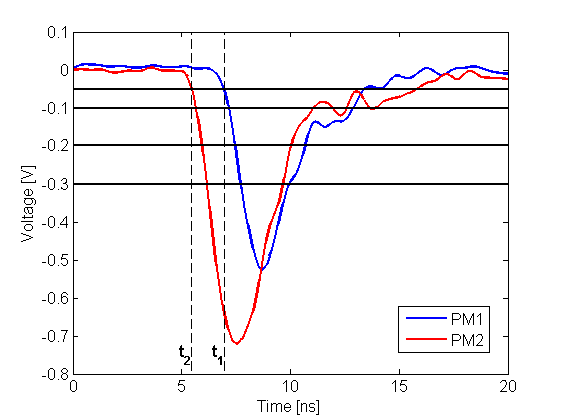}}
\caption{
Example of two signals registered in coincidence at both scintillator ends.
\label{Fig2}
}
\end{figure}

\subsection{Compressive Sensing framework}

Since the shape and amplitude of signals are predominantly related with the hit position, further improvement of the resolution may be provided by the analysis of the full time signals.  According to the theory of compressive sensing (CS) [8, 9], a signal that is sparse in some domain can be recovered based on far fewer samples than required by the Nyquist sampling theorem. In recent articles [10, 11, 12] we have proposed a novel signal recovery scheme based on the Compress Sensing method and the statistical analysis that fits to the signal processing scenario in J-PET devices. Under this theory only a recovery of a sparse or compressible signals is possible. In articles [10, 11] the sparse representation of signals was provided by the Principal Component Analysis (PCA) decomposition. We will not describe all the steps of signal processing here, but we just state that the recovery of full time signals based on eight samples is very accurate. For further details about signal recovery scheme in the J-PET framework the interested reader is referred to Ref. [10, 11, 12].

The application of CS theory enable to take an advantage from fully sampled signals and open an area for completely new algorithms for the estimation of the values of times $t_1$ and $t_2$ (see Fig. 2). In the following we will use the recovered signals to provide the signal normalization.

\subsection{Method of signal normalization }

Due to the low detection efficiency of the plastic scintillators (and interaction of gamma quanta predominantly via Compton effect), low number of photons reach the photomultipliers and the charge, as well as amplitude, of signals is subject to a large variations. However, the shapes of the signals are highly related with the position of the interaction. In order to improve the time resolution, a method of signals normalization is proposed which permits to decrease the smearing of signals charge. The procedure of signal normalization is as follows.

Consider $L$ data sets representing charges of signals at PM1 and PM2 gathered for $L$ positions. The mean values of the charges at positions along the strip ($i = 1, .., L$) at PM1, PM2 will be denoted by $Q_{m1(i)}$ and $Q_{m2(i)}$, respectively. Furthermore, the standard deviation of charges at each position along the strip ($i = 1, .., L$) at PM1, PM2 will be denoted by $Q_{s1(i)}$ and $Q_{s2(i)}$, respectively.

Suppose that new pair of signals $s_1$ and $s_2$ have been recovered based on time samples registered at four amplitudes levels at PM1 and PM2, respectively. The charges of the signals, $Q_1$ and $Q_2$, are calculated as an integrations of the $s_1$ and $s_2$ functions, respectively. The proposed normalization procedure qualifies a new measurement, represented by $Q_1$ and $Q_2$, to the $j^{th}$ position:
\begin{equation}
	j = \arg \min ((Q_{m1(i)} - Q_1)^2 + (Q_{m2(i)} - Q_2)^2).
\end{equation}
Next, the recovered signals $s_1$ and $s_2$ are normalized according to the formula:
\begin{align}
	s_{n1} = s_1 \frac{Q_1}{Q_{m1(j)}} \\
	s_{n2} = s_2 \frac{Q_2}{Q_{m2(j)}} 
\end{align}
where $s_{n1}$ and $s_{n2}$ denote the normalized signals. 

\section{EXPERIMENTAL RESULTS}

In the first step of the analysis the charge distributions of signals were investigated. Experimental results based on the signals registered along the scintillator strip are presented in Fig. 3. Mean values of charges at PM1(2) are marked with solid blue (red) curves. As expected, the curves are symmetrical with respect to the center of the scintillator strip (position of 
15~cm). The distributions of $Q_{m1}$+/-$Q_{s1}$ and $Q_{m2}$+/- $Q_{s2}$ at PM1(2) along the strip are marked with dashed blue (red) curves. From Fig. 3 one may observe that $Q_{s1}$ and $Q_{s2}$ have the same trend as $Q_{m1}$ and $Q_{m2}$, respectively, and are in the range from about 6 to 20~pC.

\begin{figure}[h!]
	\centerline{\includegraphics[width=0.7\textwidth]{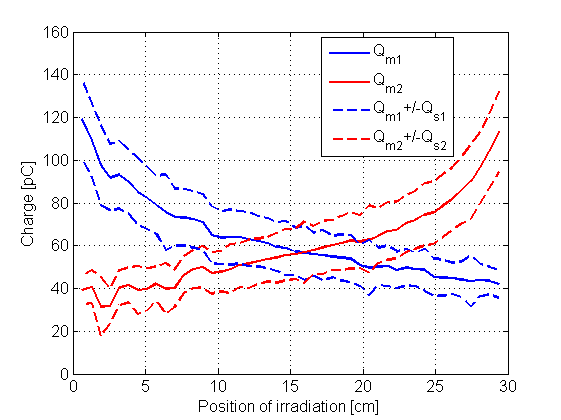}}
\caption{
Charge distributions along the scintillator strip.
\label{Fig3}
}
\end{figure}

\subsection{An example of the signal normalization}

Figure 4 shows an example of the normalization of signals registered at PM1 at fixed position 5~cm from PM1. The three corresponding signals registered at second photomultiplier (PM2) are not shown here, but were used during the normalization process to estimate the position of irradiation (see Eq. 4). The left part of Fig. 4 shows a three randomly selected raw signals registered via scope. The right part of Fig. 4 presents the same signals after the normalization procedure provided according to the description in Sec. 2.4. The same colors of the signals on the left and right part in Fig. 4 indicate a corresponding pair of signals before and after normalization. As it is seen from left part of Fig. 4 the shapes of the three signals are similar but the charges differ. However, an estimated positions according to Eq. 4 were found to be very close, and therefore the charges of the normalized signals were also very similar (see right part of Fig. 4). 

\begin{figure}[h!]
\begin{tabular}{c c}
	\includegraphics[width=0.49\textwidth]{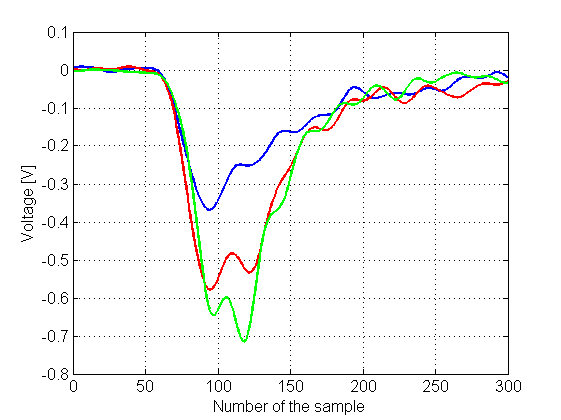} & \includegraphics[width=0.49\textwidth]{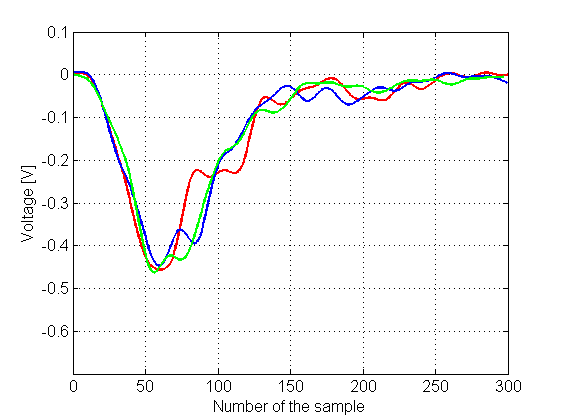}\\
\end{tabular}
\caption{
Distributions of the amplitude signals gathered at central (left panel) and left end (right panel) position of the strip. A sharp edge of the spectrum for the PM1 is due to the triggering conditions, as described in the text.
\label{Ampl_distrib}
}
\end{figure}

\subsection{Time resolution of the event time reconstruction}

The normalization method was verified using signals from all the irradiation positions. Each pair of signals was recovered via compressed sensing method based on the information from four amplitude levels (see Fig. 2) and next normalized according to the description in Sec. 2.4. In order to compare the time resolutions before and after normalization process, the normalized signals, $s_{n1}$ and $s_{n2}$, were sampled in the voltage domain at four levels from 50 to 300~mV (see Fig. 2). For each position and at each level the distribution of time difference was calculated. 

In Fig. 5 and 6 the resulting resolutions ($\Delta t$) are presented as a function of irradiated position, determined when applying the lowest threshold (50~mV) and the highest threshold (300~mV), respectively. In Fig. 5 and 6, the resolutions obtained based on the raw and normalized signals are indicated with blue and red squares, respectively. As expected, due to the time walk effect the resolution determined when applying threshold at 50~mV is better with respect to the resolution obtained at 300~mV. However, the influence of time walk effect is more visible in the case of raw signals. From Fig.~5 and 6 one can infer that the time resolution is almost independent of the position of irradiation. An average resolution of the time difference along the strip at the lowest tested amplitude level (Fig.~5) was determined to be 172~ps and 160~ps for the raw and normalized signals, respectively. This corresponds to the improvement of the resolution of the moment of the interaction from 86~ps to about 80~ps.

\begin{figure}[h!]
	\centerline{\includegraphics[width=0.7\textwidth]{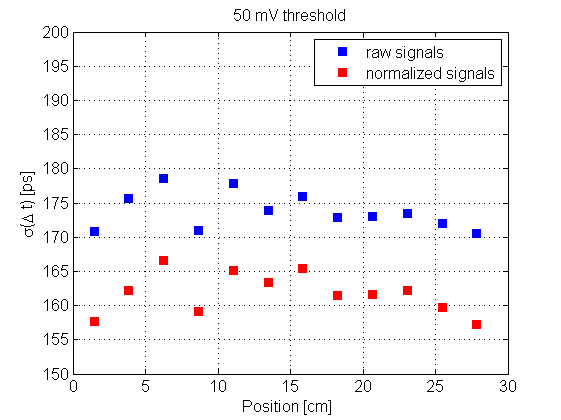}}
\caption{
Time resolution of the reconstruction after appying the lowest amplitude threshold of 50 mV.
\label{Fig5}
}
\end{figure}

\begin{figure}[h!]
	\centerline{\includegraphics[width=0.7\textwidth]{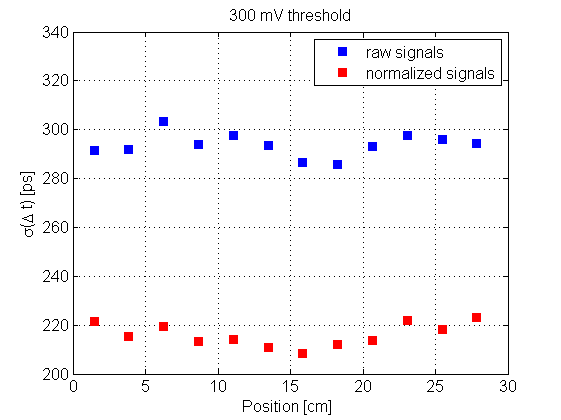}}
\caption{
Time resolution of the reconstruction after appying the highest amplitude threshold of 300 mV.
\label{Fig6}
}
\end{figure}

\section{CONCLUSIONS}

In this paper the concept of signal normalization in a novel J-PET scanner was introduced. J-PET device is based on plastic scintillators and therefore is a promising solution in view of the TOF resolution. In a related works [10, 11] it was shown that Compressive Sensing theory can be successfully applied to the problem of signal recovery in a J-PET scanner. The information from fully recovered signals was utilized in order to provide the normalization of the signals. It was shown that with fully recovered signals a better time resolution of J-PET scanner is achieved; the resolution of the moment of the interaction was improved from about 86~ps to 80~ps. It should be stressed that different approaches for utilizing the recovered information from Compressive Sensing theory may be considered and the studies are in progress.

\section{Acknowledgements}
We acknowledge technical and administrative support of T. Gucwa-Ry\'s, A. Heczko, M. Kajetanowicz, G. Konopka-Cupia\l, 
J. Majewski, W. Migda\l, A. Misiak, and the financial support by the Polish National Center for Development 
and Research through grant INNOTECH-K1/IN1/64/159174/NCBR/12, 
the Foundation for Polish Science through MPD programme, 
the EU and MSHE Grant No. POIG.02.03.00-161 00-013/09, 
Doctus - the Lesser Poland PhD Scholarship Fund. 

%\section*{References}

%
\end{document}